# Beautimeter: Harnessing GPT for Assessing Architectural and Urban Beauty based on the 15 Properties of Living Structure

Bin Jiang

LivableCityLAB, Thrust of Urban Governance and Design
The Hong Kong University of Science and Technology (Guangzhou), China
Email: binjiang@hkust-gz.edu.cn

*"Everything should be made as simple as possible, but not simpler."*
– Albert Einstein

*"Beauty is not about how it looks, but about how it is."*
– Christopher Alexander

**Abstract:**
Beautimeter is a new tool powered by generative pre-trained transformer (GPT) technology, designed to evaluate architectural and urban beauty. Rooted in Christopher Alexander's theory of centers, this work builds on the idea that all environments possess, to varying degrees, an innate sense of life. Alexander identified 15 fundamental properties, such as levels of scale and thick boundaries, that characterize living structure, which Beautimeter uses as a basis for its analysis. By integrating GPT's advanced natural language processing capabilities, Beautimeter assesses the extent to which a structure embodies these 15 properties, enabling a nuanced evaluation of architectural and urban aesthetics. Using ChatGPT, the tool helps users generate insights into the perceived beauty and coherence of spaces. We conducted a series of case studies, evaluating images of architectural and urban environments, as well as carpets, paintings, and other artifacts. The results demonstrate Beautimeter's effectiveness in analyzing aesthetic qualities across diverse contexts. Our findings suggest that by leveraging GPT technology, Beautimeter offers architects, urban planners, and designers a powerful tool to create spaces that resonate deeply with people. This paper also explores the implications of such technology for architecture and urban design, highlighting its potential to enhance both the design process and the assessment of built environments.

**Keywords**: Living structure, structural beauty, Christopher Alexander, AI in Design, human centered design

## 1. Introduction

Within the field of design, architectural and urban beauty has long been a central focus, yet few scholars have explored it as deeply as Christopher Alexander (1979, 1994, 2002–2005) and his co-workers (Alexander and Huggins 1964, Alexander and Carey 1968). Among Alexander's significant contributions is his theory of centers, which suggests that a built environment's beauty and coherence derive from the presence and arrangement of centers—distinct parts that contribute to a larger, cohesive whole. This theory posits that spaces possess an inherent 'aliveness' or 'livingness' when these centers resonate with our inner experiences, fostering a deep, intuitive connection with those who inhabit or perceive these spaces (Alexander 2002–2005). Although the theory of centers offers profound insights into architectural and urban design, its practical application presents notable challenges. The theory's reliance on subjective perception of centers makes consistent evaluation across individuals and contexts complex, and the need for nuanced interpretation limits its scalability in broader architectural and urban design contexts.

The central element of Alexander's work is the concept of living structure, or "livingness", which refers to the degree to which the qualities of a space or environment enhance life and resonate with human experience. A living structure (detailed further in Section 2 for their 15 fundamental properties) has a profound coherence, harmony, and interconnectedness, where each part makes a meaningful contribution to the overall whole. More than just being a physical characteristic, this livingness is a quality that emerges from the interplay of various elements within a space, creating a unified and vital environment. Alexander (2002–2005) contended that the existence of living structure in a built environment makes a space feel alive, inviting, and capable of nurturing human well-being (Alexander 1979, Lewicka 2011). Essentially, the theory of centers assesses the degree of living structure within a space. When individuals engage with this theory, they reflect on whether space embodies a living structure that resonates with their own sense of aliveness (Rofè 2016). This reflective process allows for a personal evaluation of urban and architectural beauty, examining how the functional and aesthetic qualities of a space align with the patterns of life that individuals intuitively recognize within themselves.

New opportunities to address these challenges have arisen because of recent advancements in artificial intelligence (AI), particularly in natural language processing. Generative pre-trained transformer (GPT) technology, which is exemplified by tools such as ChatGPT (OpenAI et al., 2023), has shown exceptional ability to understand and generate human-like texts. These technologies excel at processing and analyzing large volumes of data, including both images and texts, which makes them ideal for tasks that demand personal and introspective assessments (e.g., Fu et al. 2023, Peng et al. 2023, Ramm et al. 2024). We have built on this potential to develop a novel tool called Beautimeter, which identifies and scores the presence of the 15 fundamental properties to evaluate the living structure within spaces. Beautimeter provides a systematic way to assess urban and architectural beauty by quantifying each property, enabling a broad and consistent evaluation of how well a space aligns with the principles of living structure. This approach offers profound insights into how individuals perceive and interact with built environments by grounding evaluations in the tangible presence of these key properties.

This paper outlines the development and application of Beautimeter, exploring how GPT technology was used to create the tool, demonstrating its use via case studies, and discussing the implications of our findings for architectural and urban design and evaluation. The results of the case studies suggest that Beautimeter outperforms by far average human judgment in assessing architectural and urban beauty. In this paper, we have connected the seemingly subjective perception of centers or living structure with the capabilities of AI, contributing to the discourse on how AI technology can deepen our understanding of beauty and help create urban environments that resonate with life-enhancing qualities.

The remainder of the paper is organized as follows. We begin with a theoretical framework outlining the theory of centers and its significance in relation to architectural and urban beauty, emphasizing the concept of living structure. The next section covers the development, design, functionality, and implementation of Beautimeter. We then present case studies where Beautimeter was applied to various images, including city scenes, buildings, paintings, and carpets, along with an analysis of the results. In the discussion section, we examine the broader implications of Beautimeter and the study overall in relation to using AI to assess architectural and urban beauty, including the ethical considerations of AI usage. We conclude with a summary of key insights and reflections on Beautimeter's potential impact on the fields of architecture and urban design.

## 2. Theoretical Framework

In this section, we look at the foundational concepts underpinning the development and application of Beautimeter. We begin by introducing the concept of living structure and the 15 fundamental properties that characterize spaces with a high degree of life, followed by two early surveys conducted by Alexander (2002–2005). We aim to provide a theoretical framework that forms the foundation of Beautimeter.

### 2.1 Living Structure and the 15 Fundamental Properties

Alexander (2002–2005) conceived and developed the concept of living structure from his earlier work on pattern languages and Turkish carpets (Alexander et al. 1977, Alexander 1979, 1993). His concept of living structure extends beyond biological systems to include the structural qualities of any object or system. For example, the banyan trees shown in Figure 1 continue to embody living structures even when they are no longer biologically alive, showing that the idea of living structure continues beyond the biological realm into the realm of structural organization. Essentially, a living structure is characterized by the presence of a recurring pattern containing significantly more small substructures than large ones, which creates a rich hierarchy of forms. This trait applies whether the structure is biologically alive or not. Investigating the number of substructures in a structure, and the hierarchy that these substructures form shows the degree of livingness or life in the structure (Jiang and de Rijke 2023). Humans can intuitively perceive this living quality, or the quality without a name (QWAN) (Alexander 1979), and often associate it with a sense of beauty. Living structure is to beauty what temperature is to warmth; the greater the degree to which a structure embodies the characteristics of living structure, the more the structure resonates with the innate human sense of beauty.

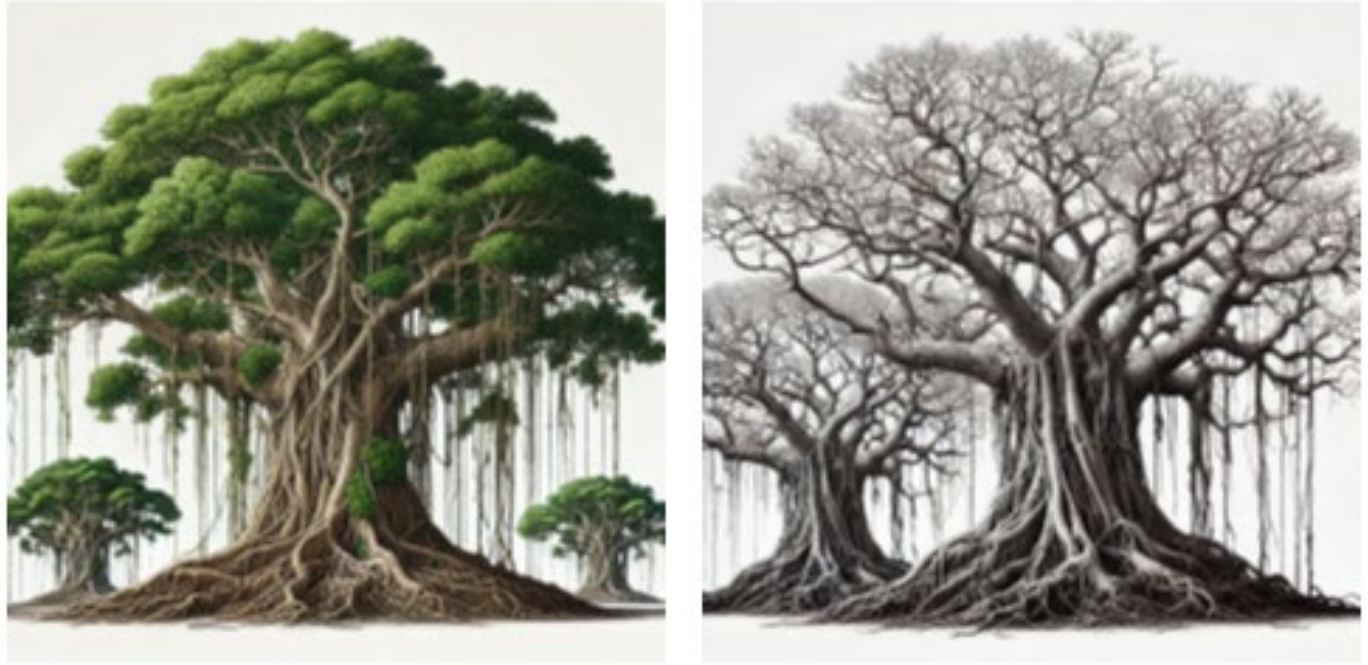

Figure 1: (Color online) Banyan trees, whether alive or dead, remain a living structure
(Note: The figure generated by Dall-E 2 suggests that the concept of living structure is a reference to its underlying form and organization rather than its biological state, alive or dead. The choice of banyan trees for this example was inspired by the concept presented in a classic paper that Alexander (1965) wrote, entitled *A City is Not a Tree*. In this context, the interconnecting branches of a banyan tree create a complex network that makes it resemble a city more closely than a 'typical' type of tree would, where two parallel branches can only connect through their parent branches.)

The 15 properties of living structure provide insights into what makes certain objects, spaces, or designs feel alive and harmonious (Figure 2). Rather than being abstract concepts, these properties are deeply rooted in how people perceive and interact with the world around them. These properties reflect the underlying order and coherence that contributes to a structure's livingness or beauty, whether the structure is a building, a natural form, or even a city. When architects, planners, and urban designers understand and apply these properties, they can create environments that resonate with the intrinsic patterns of life and foster spaces that are functional and aesthetically pleasing. Among the 15 properties, the levels of scale property refers to the presence in a structure of multiple scales of size, from the smallest detail to the largest form, which creates a sense of balance and cohesion. Strong centers highlight the significance of focal points in a design, drawing the eye and creating a sense of order. Another key property is thick boundaries, which define the edges of substructures and spaces; this helps delineate and protect the integrity of forms. Alternating repetition and gradients introduce variation and transitions within a design, making it dynamic and engaging. Properties like local symmetries and positive space also contribute to a structure's sense of fullness and order. The 15 properties combine to deliver a comprehensive framework with which to understand and create living structures, with the beauty and coherence of the design being tied to its transformational and geometric characteristics.

The 15 properties can be distilled into two overarching laws – the scaling law (Jiang 2015) and Tobler's law (Tobler 1971) – that are foundational for understanding and characterizing living structure. Firstly, the scaling law is essentially the levels of scale property, asserting that, across any given structure, a hierarchy exists in which there are far more small substructures (or centers) than large ones. This distribution is seen across several levels of scale, from the smallest components within a system to the most significant elements. Having many small centers contributes to the overall structure's richness, complexity, and coherence, enabling it to exhibit the qualities of life and vibrancy that are characteristics of a living structure. The scaling law shows the importance of having a fine-grained hierarchy in order to create environments that resonate with human experience, and many of the 15 properties (e.g., alternating repetition, contrast, roughness, positive space, and local symmetries) recur at different levels of scale.

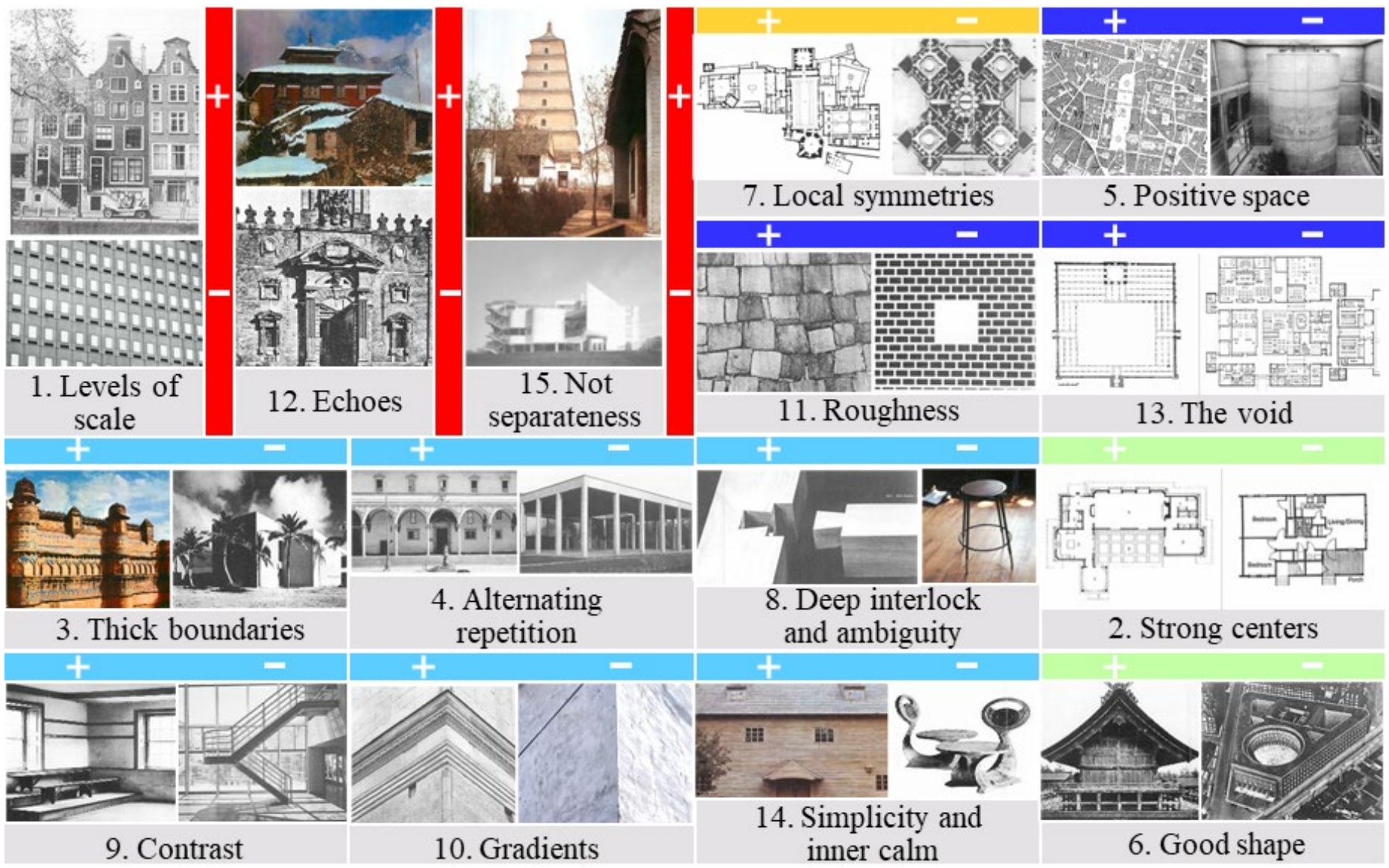

Figure 2: (Color online) Illustration of the 15 properties of living structure
(Note: Each of the 15 key properties of living structures is accompanied by positive (+) and negative (-) examples, primarily based on Alexander (2002–2005, Chapter 5 of Book 1 in particular). The properties are categorized into five groups, each of which is marked by a different color. The red group includes global properties that operate across multiple scales, while the two blue groups focus on properties at the local level. The local symmetries property is the earliest of the 15 properties, whereas the two properties, strong centers and good shape, are inherent to living structure itself, making them redundant.)

Tobler's law, which is often cited as the first law of geography, complements the scaling law by stating that the substructures within each level of scale tend to be more or less similar in terms of size (Tobler 1971). Tobler's law ensures that there is balance and harmony among the substructures at every level of a structure, adding to the environment's overall coherence and "aliveness". Together, the scaling law and Tobler's law provide a framework for understanding living structures. The former draws attention to the need for a hierarchical distribution of sizes across scales, while the latter ensures that there is a proportional and harmonious arrangement of elements within each level (e.g., alternating repetition, contrast, roughness, positive space, and local symmetries). From a living structure perspective, the scaling law should be recognized as the primary law because the hierarchical richness it describes is fundamental to the creation of living structures. Meanwhile, Tobler's law provides a secondary effect, ensuring that the relationships among the elements within each scale are balanced and coherent.

### 2.2 Two Surveys about the Mirror-of-the-Self Test (MOST)

MOST is an important part of Alexander's broader architectural and urban theory, which emphasizes the creation of environments that resonate with people at an intrinsic level. The MOST was developed as an introspective tool and designed to assess whether a space possesses what Alexander (1979, 1993) termed QWAN (Gabriel 1998); that is, a sense of life, beauty, or wholeness (Alexander 2002–2005) that evokes a deep connection among the people who inhabit or perceive it. The MOST is based on Alexander's belief that all spaces can reflect, to some degree or other, people's emotional and psychological states. Based on his own work on pattern languages (Alexander et al. 1977, Alexander 1979) and on what makes buildings and environments feel "alive", Alexander (2002–2005) argued that people are able to intuitively recognize to what degree an architectural or urban environment reflects their sense of self, not just in an aesthetic sense, but related to a deeper – often unconscious – connection with the environment.

The MOST is part of Alexander's wider theoretical framework, which centered around the concept of living structure. Alexander argued that living structures are characterized by a high degree of order and coherence and multiple interrelated centers or substructures (Gabriel and Quillien 2019, Jiang and Huang 2021). These substructures are brought to life by applying the 15 fundamental properties listed in Figure 2. Each property contributes to the overall harmony and life of a space and makes the space resonate with the human spirit. The MOST is inherently objective, although it relies on an individual's intuitive response to a space, asking individuals to state whether they view themselves as reflected in the space and whether the space feels like a natural extension of their inner world. The MOST focuses on a space's emotional and psychological impact, aiming to determine the degree to which a space or an object has the life-giving qualities that, according to Alexander, are essential to achieve a truly human-centered built environment.

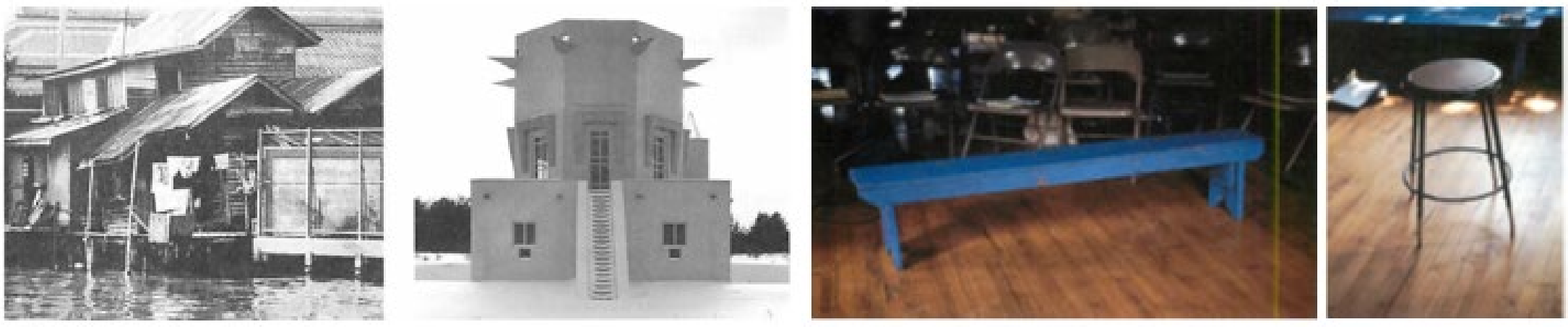

Figure 3: (Color online) Two pairs of images for the two surveys
(Note: Bangkok slum house versus a postmodern house, and a painted blue bench versus a gray steel stool.)

In 1992, Alexander (2002–2005) surveyed architecture students at the University of California, Berkeley, regarding their perceptions of life in architectural and urban designs. The students were presented with the two contrasting images shown in Figure 3 – a slum house in Bangkok and a postmodern octagonal tower – and asked which building they felt had more life. Notably, 81% of the students (89 out of 110) felt that the Bangkok slum house had more life, and none chose the postmodern tower. This overwhelming majority among students who had been trained in contemporary architectural and urban models highlighted a disconnect between their education and their instinctive perceptions and challenged prevailing architectural and urban ideals.

In 1985, at a conference in New York, Alexander (2002–2005) conducted a similar exercise, asking 100 participants to state which of two objects – a gray steel stool and a blue-painted wooden bench (see Figure 3) – better reflected their sense of self. Only one of the 100 participants chose the steel stool. While he defended his choice as subjective at first, he later changed his mind and opted for the wooden bench. Examples like this underline the profound emotional influence of Alexander's approach and reveal universal truths regarding how people intuitively relate to designs that reflect their inner sense of life and self.

**3. Development of Beautimeter**

A significant innovation in assessing architectural and urban beauty emerged with the integration of GPT technology, specifically ChatGPT (OpenAI et al. 2023), into the theory of centers. Our goal in incorporating GPT was to standardize and scale the evaluation process, enabling broader application and analysis. The natural language processing capabilities of GPT allow it to interpret and respond to prompts related to the beauty and coherence of architectural and urban images through the lens of the theory of centers and its 15 fundamental properties. The primary challenge was ensuring that GPT could navigate the nuanced and objective nature of these assessments while remaining aligned with the principles of life-enhancing design. The resulting system prompts GPT to evaluate architectural and urban images using a structured yet adaptable rubric, leading to the generation of an overall beauty score that reflects the combined influence of the 15 properties.

### 3.1 Design and Functionality

The development of Beautimeter aimed to create a user-friendly tool powered by GPT that incorporates the theory of centers and the 15 fundamental properties of beauty. The design process began by defining the core functionality: enabling GPT to analyze two architectural and urban images and to determine which one embodies a higher degree of beauty based on these properties. This was achieved by translating the properties into a scoring system that GPT could effectively utilize. By leveraging GPT's advanced natural language processing capabilities, Beautimeter prompted users to present two images of our daily lives (direct image inputs), indicating which is more beautiful and providing the corresponding score for each image.

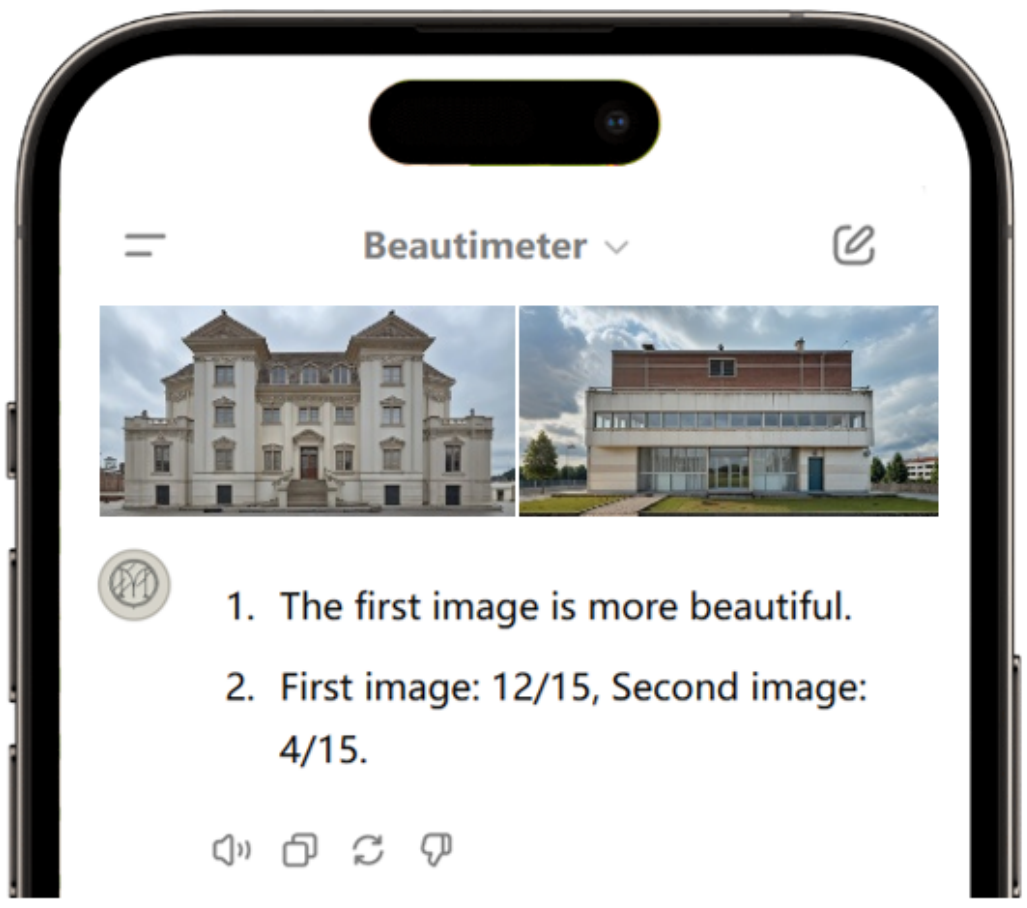

Figure 4: (Color online) Beautimeter: A GPT-powered tool for measuring beauty

Behind Beautimeter is the prompt like the following:

> *"Given the two input images, I would like to know which is more beautiful or has a higher degree of beauty according to the 15 properties of living structure defined by Christopher Alexander. Score the two images based on the 15 properties, with 15 being the highest and 0 being the lowest; in other words, each of the 15 properties is scored between 0 and 1. Please refrain from articulating or elaborating on details."*

GPT processes the prompt and compares the images based on the encoded understanding of the 15 properties or the theory of centers in general and generates an overall beauty score for each image. The user interface (UI) of Beautimeter (Figure 4) was designed to be intuitive so that users could easily upload or capture any images and obtain results. The tool presents users with the comparison task, processes their input or two images, and then clearly and concisely displays the results (Figure 4). The UI also allows users to track their evaluations and compare results over time, thus offering a deeper understanding of how different spaces resonate with the principles of life-enhancing design.

### 3.2 Implementation

The implementation of Beautimeter faced virtually few challenges, demonstrating the powerful capabilities of GPT rather than the complexities of my research. As an expert in living structure theory, I can objectively assess the effectiveness of GPT in understanding the theory of centers and the 15 properties. We found that the natural language processing capabilities of GPT were inherently sufficient to reflect the nuances of the 15 properties. We tested the tool on a variety of images, including buildings and urban environments. Notably, despite our efforts to fine-tune the model, we discovered that GPT's out-of-the-box performance was already robust enough to meet our needs.

Throughout the development process, our focus was on maintaining the integrity of Alexander's principles while leveraging GPT's strengths. The implementation was straightforward, resulting in a tool that is both technically sound and aligned with the results of the MOST conducted by Alexander (2002–2005). Ultimately, Beautimeter automates and enhances the assessment of architectural and urban beauty, offering users a profound understanding of human experience without the need for extensive adjustments.

## 4. Case Studies for Verification

We conducted a series of experiments as case studies that aimed to assess the ability of Beautimeter to evaluate and compare the aesthetic qualities of various architectural and urban images. We examined pairs of images, each of which represented varying degrees of beauty or life, as defined by Alexander (2002–2005), seeking to validate Beautimeter's effectiveness at capturing the nuanced and seemingly subjective nature of human aesthetic judgment. The sub-sections below outline the methodology and results of the experiments and present some key insights into the relationship between living structure and perceived beauty.

### 4.1 Experiments with Pairs of Images

We investigated Beautimeter's applicability by conducting a series of experiments that involved evaluating 46 pairs of images. The images were sourced from the Nature of Order Book 1 (Alexander 2002–2005) – specifically, Chapter 2 "Degrees of Life" (Figure 5) and Chapter 8 "The Mirror of the Self" (Figure 6) – which include buildings, city scenes, paintings, and artifacts. These experiments explored the living structure concept by asking GPT which image in each pair showed a stronger sense of beauty or life. In the original context, Alexander (2002–2005) provided a form of ground truth by suggesting that the left-hand image in each pair generally embodied more beauty or life than the right-hand image. Our experiment sought to test this hypothesis by scoring each image, based on the 15 properties, via Beautimeter. The aim was to quantify the seemingly subjective yet consistent nature of human perception regarding what constitutes a living structure. Beautimeter was presented pairs of images and asked to evaluate which one exhibited a higher degree of beauty or life. Beautimeter generated a score for each image based on the 15 properties. The scoring did not include any specific articulation or elaboration and instead focused solely on the overall perception of beauty, as informed by these properties.

### 4.2 Results and Discussion

Using the image pair of the Bangkok slum house and a postmodern house (Figure 3), let us first generate Table 1 to break down the beauty scores based on the 15 properties of living structure. The table provides a detailed evaluation of how each property contributes to the overall perception of beauty in both images. The traditional Bangkok slum house exhibits a higher degree of living structure, with strong centers, well-defined positive space, and a rich interplay of scales and textures. In contrast, the postmodern house, while geometrically structured, lacks integration between its ele-ments and scores lower in almost all of the properties. This breakdown example not only clarifies how GPT interprets these images but also highlights the importance of the synthesis of properties in shaping a more profound sense of beauty or life.

Table 1: Beauty score breakdown for the Bangkok slum house (left) and the post-modern house (right) based on the 15 properties

| Property | Left Image | Right Image | Property | Left Image | Right Image |
|---|---|---|---|---|---|
| Levels of Scale | 0.9 | 0.6 | Contrast | 0.85 | 0.7 |
| Strong Centers | 0.85 | 0.7 | Gradients | 0.78 | 0.6 |
| Boundaries | 0.88 | 0.65 | Roughness | 0.9 | 0.5 |
| Alternating Repetition | 0.87 | 0.6 | Echoes | 0.86 | 0.55 |
| Positive Space | 0.9 | 0.7 | The Void | 0.96 | 0.4 |
| Good Shape | 0.83 | 0.68 | Simplicity & Inner Calm | 0.88 | 0.6 |
| Local Symmetries | 0.8 | 0.7 | Not-Separateness | 0.92 | 0.4 |
| Deep Interlock and Ambiguity | 0.82 | 0.62 | SUM | 13 | 9 |

Tables 2 and 3 summarize the results of the experiments. The tables quantify the degree of beauty, as perceived through the lens of the 15 properties, which provides insights into the concept of living structure. Table 2 displays the scores for the image pairs from Figure 5, with the left- and right-hand images scored separately. Alexander proposed that the left-hand-side images generally possess more beauty or life than those on the right, yet the yellow-highlighted results in the table show instances where the right-hand-side images received higher scores than those on the left. Table 3 quantifies the results of the image pairs from Figure 6, and there are again cases where images on the right were rated higher than those on the left by Beautimeter (again highlighted in yellow). The deviations shown in these two tables underline the complexities involved in applying the 15 properties to diverse visual contexts, revealing the power but also limitations of GPT when evaluating beauty or life across different types of images.

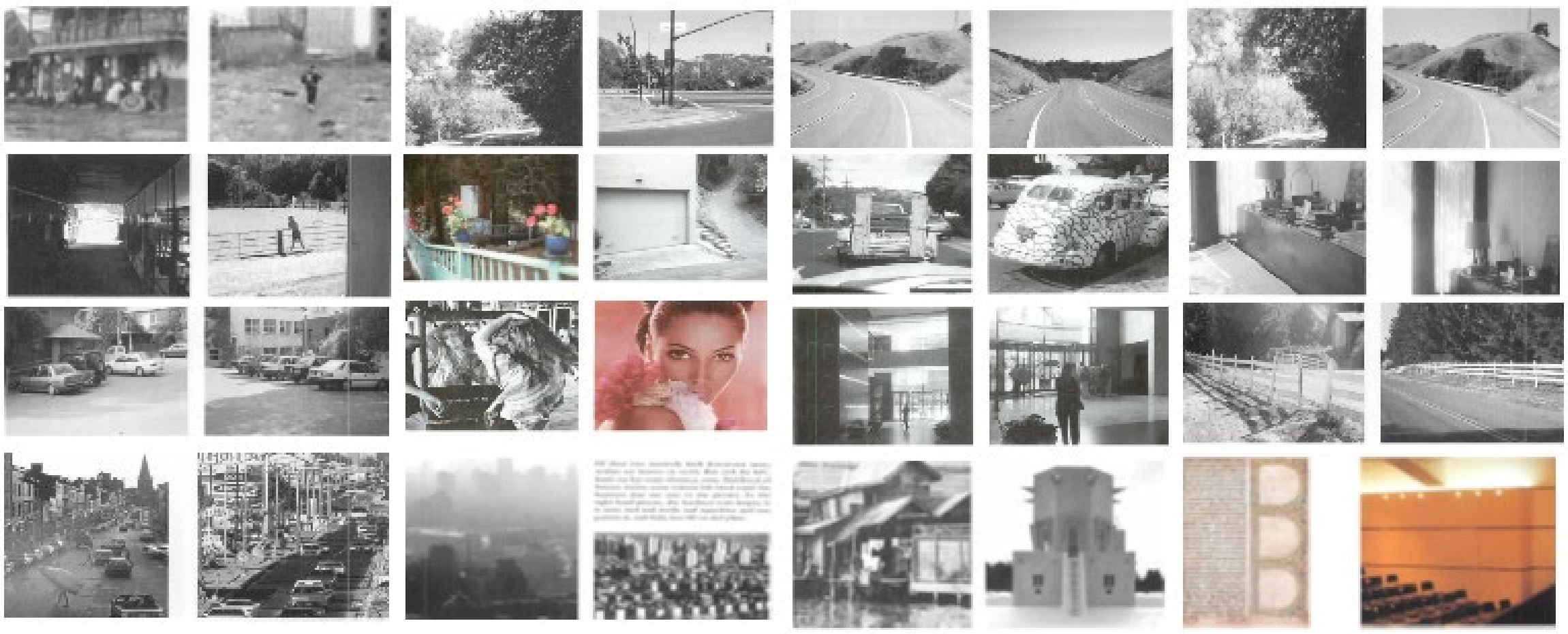

Figure 5: (Color online) Sixteen pairs of images A1–A16 numbered sequentially in reading order; for each, the left-hand image is more beautiful than the right-hand one

Table 2: Beautimeter scoring results for the image pairs shown in Figure 5

| ID | Left | Right | ID | Left | Right | ID | Left | Right | ID | Left | Right |
|---|---|---|---|---|---|---|---|---|---|---|---|
| A1 | 12.5 | 7.0 | A2 | 12.5 | 8.5 | A3 | 11.0 | 10.0 | A4 | 12.5 | 11.0 |
| A5 | 9.5 | 11.5 | A6 | 13.0 | 9.5 | A7 | 6.0 | 8.0 | A8 | 12.0 | 10.5 |
| A9 | 11.0 | 10.0 | A10 | 11.5 | 10.0 | A11 | 12.0 | 11.0 | A12 | 12.5 | 11.0 |
| A13 | 13.0 | 9.5 | A14 | 12.5 | 8.0 | A15 | 13.0 | 9.0 | A16 | 13.5 | 8.5 |

The results above offer insights into the complexity of human perception when evaluating the concept of living structure or QWAN. Most of the results align with Alexander's original assessments, where the left images score higher than the right ones. However, there are five exceptional cases where the right images outperform the left. While the reasons for this are uncertain, these cases are rare and may be considered outliers. Nevertheless, we could argue that beauty or life, as perceived in these in-stances, is not merely the sum of individual properties, but rather the integration and synthesis of various elements working together. Our findings highlight the importance of the 15 properties as a framework for assessing beauty, emphasizing the need to account for the integration of these properties rather than just their individual presence. Beautimeter, while a powerful tool for quantifying human perceptions, requires further refinement to better capture the interplay between these properties. For example, greater weight may need to be given to some properties, such as the levels of scale, echoes, and not separateness, which might be more significant than local properties. As architecture and urban design continue to evolve, tools like Beautimeter can play a vital role in ensuring that the spaces we create resonate with the deep, intuitive sense of beauty and life inherent in our environments.

Future refinements of Beautimeter could involve weighing these properties based on empirical studies or expert consensus. Previous research suggests that certain properties resonate more strongly with human perception, particularly those that contribute to a sense of coherence and wholeness. By integrating weighted scoring derived from experimental data or structured expert evaluations, the model could more accurately reflect how beauty is perceived in architecture and design. This remains an open area for further investigation and refinement.

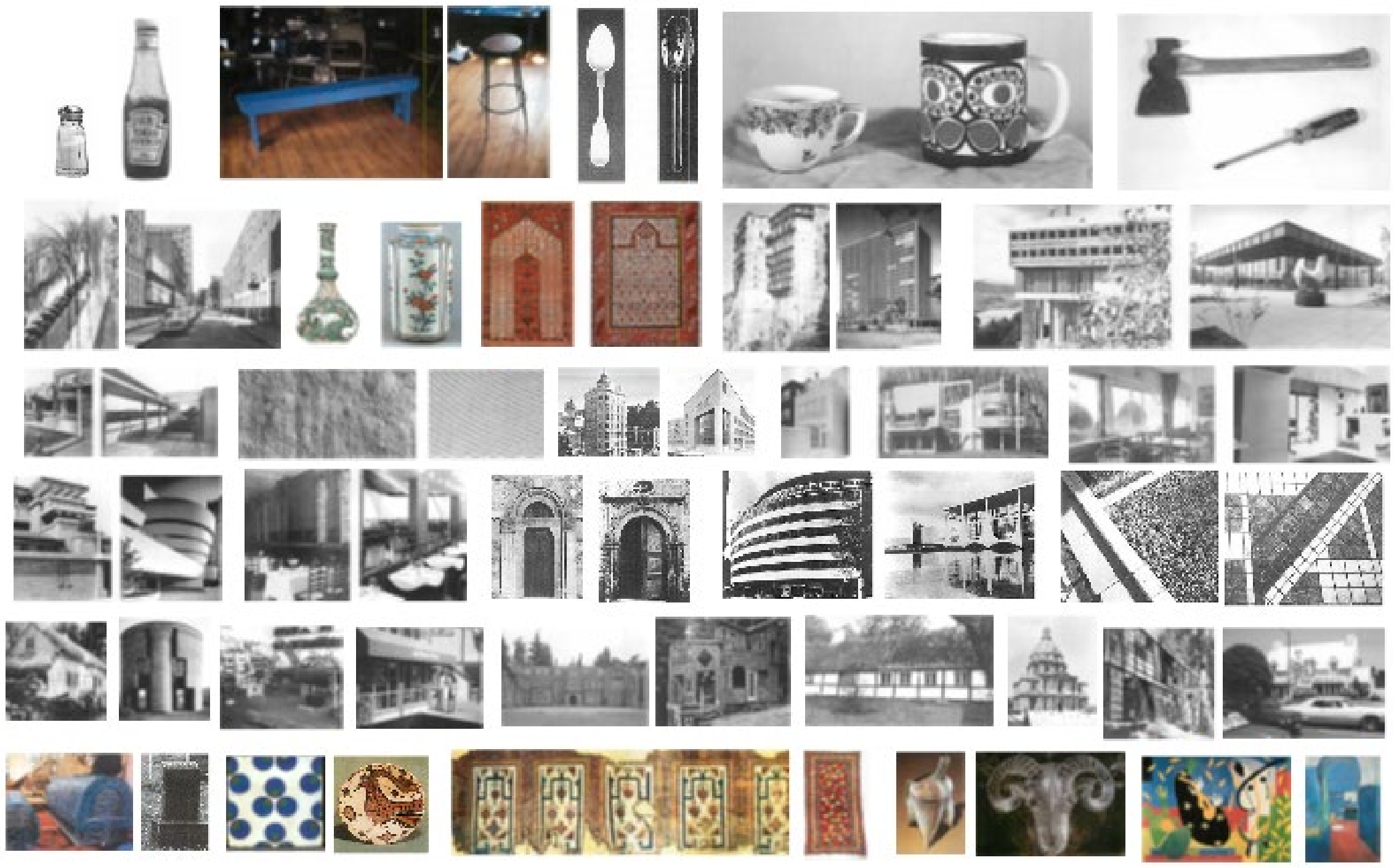

Figure 6: (Color online) Thirty pairs of images of different things numbered sequentially in reading order (in each pair, the image on the left represents something with a higher degree of life or beauty than the image on the right.)

Table 3: Beautimeter scoring results for the image pairs shown in Figure 6

| ID | Left | Right | ID | Left | Right | ID | Left | Right | ID | Left | Right | ID | Left | Right | ID | Left | Right |
|---|---|---|---|---|---|---|---|---|---|---|---|---|---|---|---|---|---|
| B1 | 7.2 | 6.5 | B2 | 7.8 | 7.1 | B3 | 8.4 | 7.6 | B4 | 8.9 | 8.1 | B5 | 9.2 | 7.3 | B6 | 11.0 | 8.4 |
| B7 | 10.8 | 9.1 | B8 | 11.3 | 9.5 | B9 | 11.7 | 7.8 | B10 | 8.9 | 7.7 | B11 | 9.1 | 8.2 | B12 | 8.4 | 7.6 |
| B13 | 10.6 | 8.8 | B14 | 10.3 | 7.9 | B15 | 9.6 | 8.3 | B16 | 10.7 | 8.9 | B17 | 9.8 | 8.4 | B18 | 10.5 | 9.2 |
| B19 | 9.7 | 8.9 | B20 | 9.4 | 8.6 | B21 | 11.0 | 8.5 | B22 | 9.2 | 8.0 | B23 | 10.9 | 9.3 | B24 | 10.5 | 9.7 |
| B25 | 9.8 | 8.4 | B26 | 10.9 | 8.7 | B27 | 9.1 | 11.3 | B28 | 10.2 | 9.4 | B29 | 9.8 | 10.7 | B30 | 8.7 | 9.3 |

While Beautimeter offers a structured approach to assessing living structure, it is not without limitations. One such case of misjudgment A5 is presented in Table 2, where the tool assigned a higher life score to the right image (an open, sunlit outdoor setting with a human figure) than to the left image (a dark, enclosed tunnel-like space). As a researcher studying Alexander's theory of living structure, I remain unconvinced by this assessment. This discrepancy may arise from either a misjudgment by Alexan-der himself or an error in GPT's evaluation. However, it is important to note that such inconsistencies are rare, occurring in fewer than 10% of cases.

Beautimeter relies mainly on the 15 properties of living structure to automatically assess and score pairs of images in terms of their livingness thanks to the advance of GPT technology. While Beautimeter is largely relying on the number of the 15 properties, other methods gauge architectural and urban beauty in a more quantitative manner (e.g., Birkhoff 1933, Palmer et al. 2013). An example is based on the formula L = S * H, where L represents the livingness or perceived beauty of a structure, S represents the number of substructures, and H denotes their hierarchical levels. This formula, which is derived from previous work (Jiang and de Rijke 2023) offers an objective and mathematical way of assessing architectural and urban beauty by analyzing the structural and hierarchical properties of spaces.

Beautimeter does not utilize the L = S * H formula directly, but the two approaches are complementary because they address different aspects of architectural and urban evaluation. Beautimeter captures the number of the 15 properties, while the L score offers a quantitative measure and focuses on the structural characteristics that contribute to the sense of life in a built environment. The two methods together offer a comprehensive toolkit that enables architects and urban designers to explore the structural integrity and also the emotional impact of the spaces they create. Such a dual perspective elaborates on what makes some environments feel more alive and beautiful than others, which can lead to more holistic and human-centered architectural and urban designs.

**5. Implications of Beautimeter and this Study**

Beautimeter represents a significant advancement in assessing and designing architectural and urban spaces, reshaping our understanding of beauty in the built environment. This section explores its philosophical and practical implications, focusing on the contrasting organic and mechanical views of space. The organic perspective fosters a holistic approach, promoting designs that enhance community engagement and sustainability, while the mechanical view often prioritizes efficiency at the cost of aesthetic and emotional considerations. This dichotomy is vital, as it influences the effectiveness and ethical implications of Beautimeter, underscoring the need for a thoughtful application that respects the complexities of urban life and the responsibilities of designers.

A central aspect of Alexander's work is the distinction between the organic and mechanical views of space and the world. The organic view (Whitehead 1929, Bohm 1980, Alexander 2002–2005) perceives space as a living entity that is strongly connected to human experience and emotions, and every element contributes to a coherent whole that resonates with life and beauty. The mechanical view (Descartes 1644, 1983), by contrast, treats space merely as a physical construct that is defined by functional and geometric properties and often lacks emotional and experiential richness. The MOST is founded on the belief that the universe and us are ultimately made of the same fundamental substance: living structure. Therefore, when we engage with a more 'living' environment or object, our own sense of self and life is enhanced.

Beautimeter can potentially have a significant impact on architectural and urban design. By providing architects and urban designers with a tool for assessing architectural and urban beauty, Beautimeter offers a new way of evaluating how well a space resonates with its users. This could lead to designs that are increasingly attuned to human needs and experiences and foster environments that function effectively, but also evoke a strong sense of place and belonging (Tuan 1977, Lewicka 2011, Seamon 2018, Mehaffy and Salingaros 2017, Salingaros 2020). Architects and urban designers could use Beautimeter during the design phases to test configurations and elements so that the final design aligns with the organic view of space. Such an approach could help avoid the creation of spaces that are technically functional but fail to resonate emotionally with their inhabitants. Beautimeter's ability to provide real-time feedback on the emotional impact of design choices could revolutionize how spaces are conceived and built. However, the use of AI in such a domain also raises certain limitations. While GPT technology can understand and process human responses, it lacks the innate human ability to fully grasp the subtleties of emotional and sensory experiences. Therefore, Beautimeter should be used as a complement to human judgment and intuition in the design process rather than as a replacement for it.

Beautimeter can have significant implications for areas of design and planning other than just architecture and urban planning and design. In fields where human experiences play a critical role, such as interior design, public space planning, and landscape architecture, the ability to assess how spaces resonate with individual people could lead to more thoughtful and user-centered designs. Moreover, as cities and public spaces become increasingly complex and diverse, tools like Beautimeter can help urban planners create environments that cater to a range of emotional and cultural needs. Planners who integrate subjective assessments into the planning process could ensure that urban spaces are not only functional but also vibrant and inclusive, fostering a sense of community and belonging.

Beautimeter is built on universal geometrical properties that capture beauty and coherence in physical spaces, informed by living structure principles that resonate across cultures. While these properties are universal, their interpretation is culturally adaptable, distinguishing them from an international style or so-called Alexander style. Instead, the Alexander style exists in its underlying living structure. The tool allows urban planners to customize the 15 properties to reflect local traditions, cultural sym-bols, and aesthetic values, ensuring that spaces are not only universally appealing but also culturally meaningful. By integrating geometry with cultural and emotional data, Beautimeter helps create environments that are both visually harmonious and deeply resonant with their users.

Using AI to assess and influence human perception of space highlights some important ethical considerations. An example is the potential for AI to oversimplify or misinterpret the complexities of human experience, which could lead to designs that may appear effective in theory but, in reality, fail to meet the deeper needs of users. There is also the risk of AI reinforcing biases in design, especially if the training data or algorithms that are used are not carefully curated to reflect diverse perspectives. The potential impact of AI on the role of architects or urban designers is another ethical consideration. There is a risk that the more prevalent role of AI tools like Beautimeter could diminish the human element in design, with decisions being increasingly driven by data rather than intuition and creativity. Therefore, it is important to find the balance between the benefits of AI-assisted design and the need to preserve the organic, human-centered approach that is at the heart of Alexander's philosophy.

## 6 Conclusion

The development and application of Beautimeter, which integrates the advanced GPT technology, represents a significant advance in assessing and understanding architectural and urban beauty, and even beyond. By harnessing the intelligent capabilities of GPT, Beautimeter offers a scalable and consistent method for evaluating whether architectural and urban spaces resonate with the intrinsic sense of life and beauty emphasized by Alexander's theory of centers. Thus, Beautimeter not only automates the evaluation of architectural and urban beauty, but also enhances its applicability and accessibility across a wide range of urban and architectural environments. Exploring Beautimeter's application via a number of case studies has shown the potential of the tool to connect human aesthetic judgment with the processing power of AI. Our results highlight some of the complexities and nuances involved in perceiving beauty, revealing the strengths but also the limitations of using AI to quantify such personal yet shared experiences. While Beautimeter successfully offers valuable insights into the living structure of spaces, it also underscores the importance of maintaining a balance between AI-driven analysis and the inherently human elements of architectural and urban design.

The implications of this study extend beyond architecture and urban design, suggesting new possibilities for using AI in urban planning, design, and other fields where human experience is central. However, while integrating AI into these traditionally human-centric domains, we must remain aware of ethical considerations and ensure that technology enhances rather than diminishes the organic, intuitive aspects of design. Moving forward, Beautimeter will serve as a tool, but also as a reminder of the ongoing dialogue between technology and the timeless principles of beauty that guide architectural and urban practice.

## Images and Data Availability Statement

Most images studied in this paper are sourced from Christopher Alexander's seminal work, *The Nature of Order* (Alexander 2002–2005). This collection serves as a foundational reference for validating the effectiveness of Beautimeter. Additionally, two images depicting banyan trees were generated using DALL-E 2. The GPT technology employed in this research is the latest iteration, ChatGPT-4o, which offers enhanced precision and versatility, making it particularly powerful for Beautimeter. All images and data have been archived and can be accessed here: XXXXX (see attached file)

## Acknowledgments

This paper was prepared with the assistance of ChatGPT-4; however, the author takes full responsibility for any errors or oversights. The author expresses deep gratitude to the Christopher Alexander & Center for Environmental Structure Archive for granting permission to use their images, with special thanks to Maggie Alexander. Additionally, the anonymous referees provided invaluable feedback that significantly enhanced the quality of this study.